\documentclass[aip,jcp,reprint,superscriptaddress,showpacs]{revtex4-1}
\usepackage[dvips]{graphicx}
\usepackage{mathrsfs} 
\begin{document}

\title{Semiclassical quantization of nonadiabatic systems with hopping periodic orbits }
\author{Mikiya Fujii}
\email{mikiya.fujii@gmail.com}
\affiliation{Department of Chemical System Engineering, School of Engineering, The University of Tokyo, Tokyo 113-8656, Japan}
\affiliation{JST, CREST, Tokyo 113-8656, Japan}
\author{Koichi Yamashita}
\affiliation{Department of Chemical System Engineering, School of Engineering, The University of Tokyo, Tokyo 113-8656, Japan}
\affiliation{JST, CREST, Tokyo 113-8656, Japan}
\date{\today}

\begin{abstract}
We present a semiclassical quantization condition, i.e., quantum--classical correspondence, for steady states of nonadiabatic systems consisting of fast and slow degrees of freedom (DOFs) by extending Gutzwiller's trace formula to a nonadiabatic form. The quantum-classical correspondence indicates that a set of primitive hopping periodic orbits, which are invariant under time evolution in the phase space of the slow DOF, should be quantized. The semiclassical quantization is then applied to a simple nonadiabatic model and accurately reproduces exact quantum energy levels.  In addition to the semiclassical quantization condition, we also discuss chaotic dynamics involved in the classical limit of nonadiabatic dynamics.
\end{abstract}

\maketitle

\section{Introduction}
The mechanism of quantization remains one of the central issues in physics and chemistry in the present day. Historically, before Dirac's canonical quantization was presented~\cite{P.A.M.Dirac1925}, quantization was performed using geometrical structures such as periodic orbits and tori, which are invariant under time evolution in phase space. Thus, geometrical structures are basic concepts used for obtaining quantum steady states in `classical' quantum theory, e.g., the Bohr~\cite{N.Bohr1913}, Bohr--Sommerfeld~\cite{A.Sommerfeld1916}, and Einstein--Brillouin--Keller~\cite{J.B.Keller1958} quantum conditions, and are therefore considered to be classical counterparts of quantum steady states. Since modern quantum mechanics has become widely accepted, quantization using geometrical structures such as periodic orbits and tori has been studied in the form of `semiclassical' quantization. In particular, semiclassical quantization of chaotic dynamics, in which regular motion sometimes coexists, is being investigated~\cite{M.C.Gutzwiller1971,M.V.Berry1981,M.V.Berry1983,M.V.Berry1984,P.Cvitanovic1989,M.V.Berry1990,E.Doron1992}. Although these semiclassical studies of the quantization mechanism consider the mechanics of systems with a single potential energy surface, note that so-called `nonadiabatic' systems also exist, which can evolve on multiple potential energy surfaces but are outside the scope of the above discussion.

The nonadiabatic dynamics means transitions of quantum population over the multiple potential energy surfaces and is one of the universal quantum phenomena in coupled systems with fast and slow degrees of freedom (DOFs), e.g., electrons and nuclei~\cite{H.NakamuraBook2002,M.Baer2006}. This mechanism has been studied since the formulation of quantum mechanics. 
Nonadiabatic transitions arise in many physical and chemical contexts, such as surface scattering~\cite{G.-J.Kroes2008}, photoisomerizations in vision~\cite{D.Polli2010}, molecular dynamics control~\cite{M.Kanno2010}, quantum computing~\cite{G.Feng2013}, and organic solar cells~\cite{H.Tamura2007,A.E.Jailaubekov2012}.
In early studies, the probabilities of nonadiabatic transitions in one-dimensional systems were established as the Landau--Zener~\cite{L.D.Landau1932,C.Zener1932} and St\"{u}ckelberg~\cite{E.C.G.Stueckelberg1932} formulae. Then, approximately 60 years after the development of these formulae, the more sophisticated and applicable Zhu--Nakamura formula~\cite{C.Zhu1992,*C.Zhu1993,*C.Zhu1994,*C.Zhu1995} was presented. In addition to these formulae, a number of theoretical frameworks~ \cite{P.Pechukas1969-1,*P.Pechukas1969-2,J.C.Tully1971,W.H.Miller1972,M.F.Herman1984-1,*M.F.Herman1984-2,*M.F.Herman1985,*M.F.Herman1995,*P.-T.Dang2011,K.Takatsuka1986,E.Duemens1992,*E.Duemens1994,T.J.Martinez1996,A.Kondorsky2004,M.Amano2005,*T.Yonehara2012,P.Oloyede2006,C.Hu2007,D.V.Shalashilin2010,*K.Saita2012,N.Ananth2010,A.Adebi2010,M.Richter2011,B.F.E.Curchod2013} and numerical packages~\cite{MOLPRO,Newton-x,PYXAID} for investigating nonadiabatic dynamics in realistic and atomistic models have also been presented. 

Although investigation of nonadiabatic dynamics in realistic and atomistic models is becoming possible, the quantization mechanism (i.e., the quantum--classical correspondence) in nonadiabatic systems has not yet been fully determined. However, identifying classical counterparts of nonadiabatic transitions will promote further understanding of quantum mechanics, because quantum mechanics should be constructed on the correspondence principle. That is, quantum mechanics must conform to classical mechanics at the limit of the infinitesimal Planck's compared to action integrals. To reveal the quantization mechanism of nonadiabatic systems, some semiclassical studies have been conducted, such as those of Miller {\it et al.}~\cite{H.-D.Meyer1979,X.Sun1997}, who re-quantized a classical electron analog model obtained through Ehrenfest treatments, and Stock and Thoss~\cite{G.Stock1997}, who re-quantized a classical analog of a continuous fast quantum DOF mapped from a discrete fast quantum DOF. In both these studies the analogical fast and slow DOFs were quantized on an equal footing. So, although the nonadiabatic transitions were well reproduced semiclassically, explicit classical counterparts of the nonadiabatic transitions between the adiabatic surfaces were not found, because classical dynamics on an adiabatic surface and transitions between adiabatic surfaces were not used explicitly. 

In this paper, we propose a new semiclassical formulation to describe the quantization mechanism of nonadiabatic systems. In contrast to the previous studies, we focus explicitly on the quantization of the nonadiabatic dynamics of a slow DOF evolving on an adiabatic surface and hopping between adiabatic surfaces. Therefore, a classical counterpart expressing the physical concept of a nonadiabatic transition between adiabatic surfaces is determined. This approach can be implemented based on recent progress regarding the path integral for nonadiabatic phenomena~\cite{V.Krishna2007,J.R.Schmidt2007,M.Fujii2011}. In particular, semiclassical treatments that are based on the nonadiabatic path integral and which have been presented in Ref.~\onlinecite{M.Fujii2011} play crucial roles.

The present paper is organized as follows: Sec.~\ref{sec:theory} describes the theoretical framework in detail. As the most important argument in the present paper, a nonadiabatic trace formula based on the nonadiabatic path integral is presented. This nonadiabatic trace formula can reveal a semiclassical quantization condition for nonadiabatic systems. In addition, a new derivation of the nonadiabatic Schr\"{o}dinger equation from the nonadiabatic path integral is also presented in this section. In Sec.~\ref{sec:ex}, the semiclassical quantization condition revealed here is applied to a one-dimensional nonadiabatic model. In the course of this application, symbolic dynamics is introduced to allow the semiclassical quantization to be conducted analytically. The chaotic motion that appears at the classical limit of nonadiabatic dynamics is also discussed in this section. Section~\ref{sec:concluding} presents some concluding remarks.

\section{Theory\label{sec:theory}}
\subsection{Adiabatic semiclassical quantization: Gutzwiller's trace formula}
We begin by considering a one-dimensional adiabatic system for which Gutzwiller's trace formula~\cite{M.C.Gutzwiller1971}, based on the semiclassical representation of the density of states (DOS), $\Omega(E)$, offers a semiclassical quantization condition
\begin{eqnarray}
\Omega(E) 
&\propto& \sum_{\lambda \in {\rm PPOs}} \sum_{k=0}^{\infty} \mathscr{G}_{\lambda}^k,
\label{eq:AdiTrace01}  \\
\mathscr{G}_{\lambda} &=& \exp \left[ \frac{i}{\hbar} \left( S^{\rm cl}_{\lambda} -\frac{\hbar\pi}{2} \nu_{\lambda} \right)  \right].
\label{eq:AdiTrace02} 
\end{eqnarray}
Here, only the component that causes divergence of the DOS at the quantum energy levels is shown. This semiclassical DOS is an expansion with primitive periodic orbits (PPO), which are not repeated cycles of other periodic orbits (PO) and are indexed by $\lambda$. The classical action integral along the PPO, $S^{\rm cl}$, is defined as
\begin{eqnarray}
S^{\rm cl} = \oint dt P\frac{dR}{dt} - H(P,R) + E = \oint dR P,
\label{eq:AdiTrace03} 
\end{eqnarray}
where $R$, $P$, and $H(P,R)$ are the coordinates, momenta, and Hamiltonian of the system under consideration, respectively. Thus, $S^{\rm cl}$ is the phase space area surrounded by each PPO. The Maslov index, $\nu$, denotes the number of times the momentum is 0 during one cycle, i.e., the number of times the PPO intersects the $R$-axis in phase space. Therefore, $\mathscr{G}$ is a geometrical quantity depending on the PPO. The summation over $k$ in Eq.~(\ref{eq:AdiTrace01}) atakes account of $k$-cycles of each PPO. Considering that only one PPO exists at each energy in a one-dimensional adiabatic system, we can rewrite the DOS as 
\begin{eqnarray}
\Omega(E) \propto \left(1-\mathscr{G}\right)^{-1},
\label{eq:AdiTrace04} 
\end{eqnarray}
 and the quantization condition of an adiabatic system is, therefore, 
\begin{eqnarray}
\mathscr{G}=1.
\label{eq:AdiTrace05} 
\end{eqnarray}
Note that, for the case of a quantum harmonic oscillator, i.e., $H(p,q)={p^2}/{2m} + m\omega^2q^2/2$, with a phase space area of ${2\pi E}/{\omega}$ and $\nu$ = 2, the quantization condition is $\mathscr{G} = \exp \left[ \frac{i}{\hbar} \left( \frac{2\pi E}{\omega} -\hbar\pi\right)  \right] =1$; that is, we recover the well-known $E_n = \hbar\omega \left(n+\frac{1}{2}\right)$ energy levels.

\subsection{Nonadiabatic path integral with overlap integrals}

\begin{figure}
\includegraphics[width=\linewidth]{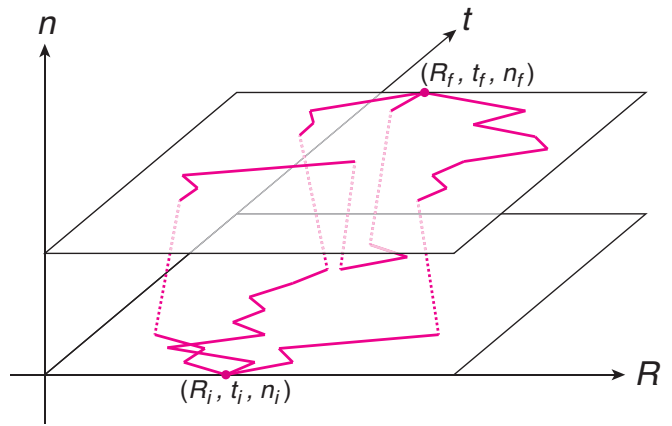}
\caption{(color online). Schematic view of the nonadiabatic path integral. Three hopping paths that begin at $(R_i, t_i, n_i)$ and end at $(R_f, t_f, n_f)$ are depicted. Each hopping path consists of red solid and dotted lines which represent the time evolution on an adiabatic surface and the transitions between adiabatic surfaces, respectively.
\label{fig:napi}}
\end{figure}
We present the semiclassical quantization condition for nonadiabatic systems in a similar manner to that used for adiabatic systems. To formulate the nonadiabatic trace formula, a nonadiabatic path integral~\cite{V.Krishna2007,J.R.Schmidt2007,M.Fujii2011} and its semiclassical approximation~\cite{M.Fujii2011} are required. 
In this subsection, we present a brief introduction to the nonadiabatic path integral. Then, the semiclassical treatment is presented in the next subsection. Here, we begin our discussion of the nonadiabatic path integral using the Born-Oppenheimer-type state ket for an arbitrary state of the total system as  
\begin{eqnarray}
| \Psi(t) \rangle = \int dR \sum_{n} \chi_{n}(R,t) |R\rangle | n; R\rangle,
\label{eq:NAPI01}
\end{eqnarray}
where $|R\rangle$ is the position ket of the slow DOF, $\chi_{n}(R,t)$ is the wave packet (probability amplitude) of the slow DOF on the $n$-th adiabatic surface, and $| n;R \rangle$ is the $n$-th eigenstate of the Hamiltonian of the fast DOF, $\hat{H}_e(R)$. Then,
\begin{eqnarray}
\hat{H}_e(R) | n; R\rangle = V_n(R) | n; R\rangle,
\label{eq:NAPI02}
\end{eqnarray}
where $V_n(R)$ is the $n$-th eigenvalue. Note that these eigenvalues and eigenstates are adiabatic energies and states that depend parametrically on the coordinates of the slow DOF, $R$. 

The nonadiabatic propagation of the probability amplitude with infinitesimal time width, $\epsilon$, is described by coordinate representation of the kernel, such that
\begin{eqnarray}
\chi_{n'}(R',\epsilon) = \sum_{n} \int dR\ K(R', n', \epsilon; R, n, 0) \chi_{n}(R,0)
\label{eq:NAPI02.5}
\end{eqnarray}
with
\begin{eqnarray}
K(R', n', \epsilon; R, n, 0) \equiv
\langle n'; R' | \langle R' | e^{-\frac{i}{h}\hat{H}\epsilon} | R \rangle  |n; R \rangle,
\label{eq:NAPI03}
\end{eqnarray}
where the Hamiltonian, $\hat{H}$, is that of the total system, which consists of the kinetic energy of the slow DOF and the Hamiltonian of the fast DOF. Therefore,
\begin{eqnarray}
\hat{H} = \hat{T}_N + \hat{H}_e(R),
\label{eq:NAPI04}
\end{eqnarray}
where $\hat{T}_N$ is the kinetic energy of the slow DOF and $\hat{H}_e(R)$ is the Hamiltonian of the fast DOF.
Applying the first order Trotter decomposition to Eq.~(\ref{eq:NAPI03}) leads to 
\begin{eqnarray}
&&K(R', \epsilon, n'; R, 0, n), \nonumber \\
&\simeq& \langle n'; R' | \langle R' | e^{-\frac{i}{h} \hat{T}_N} | R \rangle |n; R  \rangle  e^{ -\frac{i}{h}V_{n}(R) \epsilon}, \nonumber \\
&=&\langle n'; R' |n; R  \rangle \langle R' | e^{-\frac{i}{h} \hat{T}_N} | R \rangle   e^{ -\frac{i}{h}V_{n}(R) \epsilon}, \nonumber \\
&\simeq& \langle n'; R' |n; R \rangle \langle R' | e^{-\frac{i}{h} \left( \hat{T}_N + V_{n}(\hat{R}) \right) \epsilon} | R \rangle.
\label{eq:NAPI05}
\end{eqnarray}
Here, note that  $\langle R' | e^{-\frac{i}{h} \hat{T}_N} | R \rangle$ can be exactly moved to pass through $|n; R \rangle$, because the coordinate representation of the kinetic energy operator is the second derivative of the delta function\cite{J.J.SakuraiBook1985} as 
\begin{eqnarray}
\langle R' | e^{-\frac{i}{h} \hat{T}_N} | R \rangle 
&=&  \frac{-\hbar^2}{2M}\frac{d^2}{dR'^2} \delta(R'-R), \nonumber \\
&=& \frac{-\hbar^2}{2M}\frac{d^2}{dR^2} \delta(R-R'),
\label{eq:NAPI06}
\end{eqnarray}
where $M$ is the mass of the slow DOF. In Eq~(\ref{eq:NAPI05}), the overlap integral of the eigenstates of the fast DOF between different slow coordinates, $\langle n';R' | n;R \rangle$, is a key factor in the nonadiabatic transitions, while the exponential term is the usual adiabatic kernel. The time evolution of the fast DOF is exactly adiabatic if the overlap integrals of the same eigenstate are equal to 1 when the slow DOF changes from $R$ to $R'$. In contrast to the adiabatic case, the probability amplitude is transferred nonadiabatically to different eigenstates if the overlap integrals between different eigenstates are not zero.  That is, non-zero values of the overlap integrals cause a breakdown of the adiabatic dynamics.  

Repeating the infinitesimal time kernel leads to a nonadiabatic kernel for finite time width. This nonadiabatic kernel can be expressed in terms of the overlap integrals and trajectories that experience hopping between adiabatic surfaces as
\begin{eqnarray}
&& K(R_f, t_f, n_f; R_i, t_i, n_i) \nonumber \\
&\equiv& \langle n_f ;R_f|\langle R_f | \exp\left[ \frac{i}{\hbar} \hat{H}\left(t_f-t_i\right)\right] |R_i\rangle |n_i,R_i\rangle, \nonumber \\
&=&  \int \mathscr{D}\left[ R(\tau), n(\tau) \right] \xi  \exp\left[ \frac{i}{\hbar} S \right],
\label{eq:NAPI07} 
\end{eqnarray}
where $\mathscr{D}$ denotes the path integral of all paths that start from $R_i$ on the $n_i$-th adiabatic surface at time $t_i$, hop between adiabatic surfaces, and end at $R_f$ on the $n_f$-th adiabatic surface at time $t_f$. 
The action integral, $S$, and the infinite product of the overlap integrals, $\xi$, are defined along each hopping path, such that
\begin{eqnarray}
S &\equiv& \lim_{J \rightarrow \infty} \sum_{j=0}^{J-1} \left[ \frac{M}{2} \left( \frac{R(t_{j+1}) -R(t_j)}{\Delta t}\right)^2 -V_{n(t_{j})}(R(t_j)) \right] \Delta t ,
\label{eq:NAPI07.5} 
\end{eqnarray}
and 
\begin{eqnarray}
\xi &\equiv& \lim_{J \rightarrow \infty}\prod_{j=0}^{J-1} \langle n(t_{j+1});R (t_{j+1})| n(t_{j});R(t_{j}) \rangle,
\label{eq:NAPI08} 
\end{eqnarray}
respectively, where $t_j \equiv j\Delta t  + t_i$ with $\Delta t=(t_f-t_i)/{J}$.  A schematic view of the nonadiabatic path integral is shown in Fig.~\ref{fig:napi} with three hopping paths beginning at $(R_i, t_i, n_i)$ and ending at $(R_f, t_f, n_f)$. The hopping path in Fig.~\ref{fig:napi} consists of red solid and dotted lines, which represent the time evolution on an adiabatic surface and the transitions between adiabatic surfaces, respectively. 

In addition to and independently of the derivation of the nonadiabatic path integral from the Schr\"{o}dinger equation via the time propagator, 
we newly present a derivation of the Schr\"{o}dinger equation from the nonadiabatic path integral. In the nonadiabatic path integral, the infinitesimal time propagation of the probability amplitude is written as
\begin{widetext}
\begin{eqnarray}
\chi_n(R, t+\epsilon) 
= \sum_{m}\int_{-\infty}^{\infty} d\eta A  \langle n; R | m; R+\eta\rangle
   \exp\left[  \frac{i}{\hbar} \frac{M\eta^2}{2 \epsilon} - \frac{i}{\hbar} V_m(R+\eta)\epsilon   \right]\chi_m(R+\eta, t),
\label{eq:NAPI10}
\end{eqnarray}
\end{widetext}
where $\eta$ is the displacement from $R$ and $A$ is a normalization constant, such that $A = (2\pi i \hbar \epsilon/M)^{-1/2}$. Considering the first term of the exponential power, the main contribution to the integral is from the range in which $M\eta^2/2\hbar\epsilon$ varies by approximately 1 radian, i.e., $-\sqrt{2\hbar \epsilon/M} < \eta < \sqrt{2\hbar \epsilon/M}$. This is because the phase factor varies rapidly with changing $\eta$ in the case of infinitesimal time, $\epsilon$. Therefore, we expand Eq.~(\ref{eq:NAPI10}) up to the first order of $\epsilon$ and second order of $\eta$ to obtain
\begin{widetext}
\begin{eqnarray}
\chi_n(R, t+\epsilon) &=& 
	\sum_{m}\int_{-\infty}^{\infty} d\eta A 
    \exp\left[  \frac{-M\eta^2}{2 i \hbar \epsilon}\right]
    \left\{
    	\langle n; R | m; R\rangle \chi_m(R, t)
    	+\frac{1}{i\hbar} \langle n; R | m; R\rangle V_m(R)\chi_m(R, t)\epsilon \right.\nonumber \\
&&  \hspace{117pt}
    \left. 
    	+\langle n; R | m; R\rangle \frac{\partial \chi_m}{\partial R} \eta
    	+X_{nm}(R) \chi_m(R, t) \eta 
    \right. \nonumber \\
&&  \hspace{117pt}
    \left. 
    	+\langle n; R | m; R\rangle \frac{\partial^2 \chi_m}{\partial R^2} \frac{\eta^2}{2}
    	+X_{nm}(R)   \frac{\partial \chi_m}{\partial R} \eta^2
    	+Y_{nm}(R) \chi_m(R, t) \frac{\eta^2}{2} 
    \right\}, 
\label{eq:NAPI11}
\end{eqnarray}
\end{widetext}
where $X_{nm}(R)$ and $Y_{nm}(R)$ are the well-known nonadiabatic couplings
\begin{eqnarray}
X_{nm}(R) &=& \int dr \Phi_n^{*}(r;R) \frac{\partial}{\partial R}  \Phi_m(r;R),\ 
\label{eq:NAPI12} 
\end{eqnarray}
and
\begin{eqnarray}
Y_{nm}(R) &=& \int dr \Phi_n^{*}(r;R) \frac{\partial^2}{\partial R^2}  \Phi_m(r;R).
\label{eq:NAPI12.5}
\end{eqnarray}
The wave function, $\Phi_n(r;R)$, is a coordinate representation of the $n$-th eigenstate of the Hamiltonian of the fast DOF as
\begin{eqnarray}
\Phi_n(r;R) = \langle r| n;R\rangle.
\label{eq:NAPI13}
\end{eqnarray}
Finally, solving the Gaussian integrals related to $\eta$ in Eq.~(\ref{eq:NAPI11}) and considering the orthonormal eigenstate conditions $\langle n; R | m; R\rangle = \delta_{nm}$, we obtain the Schr\"{o}dinger equation with nonadiabatic couplings as
\begin{widetext}
\begin{eqnarray}
i\hbar \dot{\chi}_n(R, t) 
&=& 
	\left[
		\frac{-\hbar^2 }{2M}\frac{\partial^2 }{\partial R^2}
		+V_n(R)
	\right]
	\chi_n(R, t)  
	-
	\sum_{m} 
	\left[
    	\frac{\hbar^2 }{M}  X_{nm}(R) \chi'_m(R, t)
    	+\frac{\hbar^2 }{2M}  Y_{nm}(R) \chi_m(R, t) 
	\right] .
\label{eq:NAPI14}
\end{eqnarray}
\end{widetext}
This derivation of the nonadiabatic Schr\"{o}dinger equation is a straightforward extension of that of the adiabatic Schr\"{o}dinger equation, which was presented by Feynman\cite{FeynmanBook1965}. Further, this confirmation that the nonadiabatic Schr\"{o}dinger equation and nonadiabatic path integral can be independently derived from each other should be sufficient evidence to convince readers of the equivalence of the nonadiabatic Schr\"{o}dinger equation and the nonadiabatic path integral with overlap integrals.

\subsection{Nonadiabatic semiclassical kernel with overlap integrals}

Semiclassical treatments of the nonadiabatic path integral with overlap integrals are briefly explained in this subsection and were first derived in Ref.~\onlinecite{M.Fujii2011}.  The nonadiabatic semiclassical kernel is derived by applying stationary phase approximations (SPAs) in two steps. The first step is the application of a SPA to the path integral on an adiabatic surface between fixed hopping points, i.e., the semiclassical approximation on the adiabatic surface between the fixed hopping points is applied. The stationary phase condition (SPC) of the first SPA requires classical trajectories on the adiabatic surface. The second step involves the application of SPAs to the integrals related to the hopping points, and the SPC of the second SPA requires that the trajectories conserve momentum before and after hopping. Ultimately, applications of these stepwise SPAs to Eq.~(\ref{eq:NAPI07}) lead to the nonadiabatic semiclassical kernel with overlap integrals (NASCO)\footnote{A schematic view of these stepwise SPAs is shown in Fig. 1 of Ref.~\onlinecite{M.Fujii2011}.} as
\begin{eqnarray}
 K_{sc} 
= \left( 2\pi i \hbar \right)^{-1/2} \sum_{\lambda}  \xi_{\lambda} \left| \frac{\partial R_f}{\partial P_i} \right|_{R_i}^{-\frac{1}{2}} \exp\left[ \frac{i}{\hbar} S^{cl}_{\lambda} - i\frac{\pi}{2}\nu_{\lambda} \right].
\label{eq:NonAdiPath05} 
\end{eqnarray}
There are two differences between the nonadiabatic semiclassical kernel with overlap integrals and the conventional adiabatic semiclassical kernel~\cite{L.S.SchulmanBook2005}. The first is that all the variables in Eq.~(\ref{eq:NonAdiPath05}) are calculated along a trajectory that experiences momentum-conserving hopping between adiabatic surfaces. After one such hop and before the next, the trajectory evolves on the adiabatic surface according to the first derivative of the adiabatic surface. The second difference is that $\xi$ exists as a component of the prefactor of the exponential term. In addition, it should be noted that there are multiple hopping trajectories that have identical starting (and/or ending) phase space points (thus necessitating the sum over $\lambda$). Note that, in Ref.~\onlinecite{M.Fujii2011}, the nonadiabatic Herman-Kluk kernel with overlap integrals (NAHKO) was also presented, based on NASCO.

Some remarks concerning similarities and differences with similar semiclassical kernels for nonadiabatic systems are mentioned below. The first nonadiabatic semiclassical kernel was presented by Pechukas\cite{P.Pechukas1969-1,*P.Pechukas1969-2}, who used time-dependent electronic functions and applied SPA to the entire time range at once. Then, the semiclassical trajectories and time-dependent effective potentials are mutually referring. In NASCO, the mutual referring was resolved using time-independent electronic bases (electronic adiabatic states) and the stepwise SPAs. Herman {\it et al.} also developed a nonadiabatic semiclassical kernel, which is a solution of the time-dependent Schr\"{o}dinger equation, by adding terms calculated along hopping trajectories to the adiabatic kernel~\cite{M.F.Herman1984-1,*M.F.Herman1984-2,*M.F.Herman1985,*M.F.Herman1995,*P.-T.Dang2011}, while Pechukas's kernel and NASCO were identified as main contributing terms to the full quantum kernel through SPAs. In addition, Herman's kernel employs a historically-used assumption that each semiclassical hopping trajectory conserves its energy at hopping. This assumption, first introduced in Tully's surface hopping\cite{J.C.Tully1971}, is meaningful when nuclei are treated as classical particles. In semiclassical theories, however, nuclei are treated as quantum particles even though classical trajectories are used to construct the kernel. This assumption, therefore, has no physical meaning in semiclassical studies. Thus, energy does not need to be conserved for each semiclassical hopping trajectory for a nonadiabatic system, although the expectation value of the energy, which is calculated from the wave packet, should be conserved. In NASCO and NAHKO, conservation of the momentum instead of the energy is required as the SPC of the integral related to the hopping point. 

It should also be noted that NASCO differs from Pechukas's and Herman's kernels regarding treatment of the Berry phase. As first shown by Kuratsuji and Iida, the phase of the infinite product of the overlap integrals coincides with the Berry phase for cyclic evolution, $C$, on an adiabatic surface~\cite{H.Kuratsuji1985,I.Ryb2004}. Thus,
\begin{eqnarray}
\lim_{\Delta R\rightarrow 0}\prod_{j=1}^{J} \langle n;R_j| n;R_{j-1} \rangle 
&\simeq& \lim_{\Delta R\rightarrow 0}\prod_{j=1}^{J} \left(1- X_{nn}(R) \Delta R\right), \nonumber \\
&\simeq& \lim_{\Delta R\rightarrow 0}\prod_{j=1}^{J} \exp{\left[- X_{nn}(R) \Delta R\right]}, \nonumber \\
&=& \exp{\left[i \oint_C dR i X_{nn}(R) \right]}, \nonumber \\
&=& \exp{\left[i \Gamma(C) \right]},
\label{eq:NonAdiPath06} 
\end{eqnarray}
with
\begin{eqnarray}
\Gamma(C) \equiv \oint_C dR i X_{nn}(R). 
\label{eq:NonAdiPath06} 
\end{eqnarray}
This $\Gamma(C)$ is the Berry phase. Therefore, the Berry phase is explicitly taken into account in NASCO, as a component of the prefactor, while Pechukas's and Herman's kernels do not hold explicitly. Because the Berry phase is not concerned with the quantum effects of the slow DOFs in principle, positioning the Berry phase external to the semiclassical approximations as a component of the prefactor is considered to be appropriate.

Concerning the physical reality of semiclassical hopping trajectories, some researchers might regard the hopping trajectories as artificial and convenient trajectories for the calculation of the branching ratio in the energy domain, because sudden changes of electronic states along the hopping trajectories seem to be unphysical. In contrast to the surface hopping, mean field dynamics, which are governed by Hellman-Feynman forces with smoothly time-dependent electronic states, are sometimes preferred. This mean field dynamics avoids such sudden changes of electronic states, although unphysical dynamics on the averaged potential for trajectories after passing through an avoided crossing have also been known. Miller {\it et al.}\cite{H.-D.Meyer1979,X.Sun1997}, Duemens {\it et al.}\cite{E.Duemens1992,*E.Duemens1994}, Takatsuka {\it et al.}\cite{M.Amano2005,*T.Yonehara2012}, and Shalashilin {\it et al.}\cite{D.V.Shalashilin2010,*K.Saita2012} extensively studied the coupled dynamics of electrons and nuclei in the context of the mean field dynamics.  However, since the question {\it ``Which is physically realistic, surface hopping or mean field?"} has not been given any clear answer, we refer to a study conducted from another standpoint. A decade ago, Tanaka pointed out that a semiclassical trajectory identified using SPA can be regarded as a time series of weak values~\cite{A.Tanaka2002}. Therefore, when we remember that semiclassical hopping trajectories can be identified using SPAs, the semiclassical hopping trajectory has the potential to become physically realistic through successive weak measurements. Regardless, further discussion to clarify the physical realism of {\it ``nonadiabatic trajectories"} should be conducted from both experimental and theoretical standpoints in the future.

\subsection{Nonadiabatic trace formula}
Based on the nonadiabatic semiclassical kernel with overlap integrals (NASCO), the extension of Gutzwiller's trace formula to nonadiabatic systems is fairly straightforward as
\begin{eqnarray}
\Omega(E) 
&\propto & \sum_{\lambda \in {\rm PHPOs}} \sum_{k=0}^{\infty} \mathscr{G}_{\lambda}^k
\label{eq:NonAdiTrace01}, \\
\mathscr{G}_{\lambda} &=&  \xi_{\lambda} \exp \left[ \frac{i}{\hbar} \left( S^{\rm cl}_{\lambda} -\frac{\hbar\pi}{2} \nu_{\lambda} \right)  \right].
\label{eq:NonAdiTrace01.3} 
\end{eqnarray}
This semiclassical DOS is an expansion with primitive hopping periodic orbits (PHPO), which are not repeated cycles of other hopping periodic orbits (HPO). Therefore, the classical action integral, $S^{\rm cl}$, and the infinite product of the overlap integrals, $\xi$, are calculated along the PHPOs.

Taking the summation over $k$ in Eq.(\ref{eq:NonAdiTrace01}), similar to the adiabatic trace formula approach, we find
\begin{eqnarray}
\Omega(E) 
&\propto & \sum_{\lambda \in {\rm PHPOs}} (1-\mathscr{G}_{\lambda})^{-1}.
\label{eq:NonAdiTrace01.4} 
\end{eqnarray}
If a PHPO could exist such that its $\mathscr{G}_{\lambda}$ were equal to 1, the DOS in Eq.~(\ref{eq:NonAdiTrace01.4}) would diverge. However, this is not the case, because $\xi_{\lambda}$ is generally less than 1. Thus, the semiclassical quantization of nonadiabatic systems cannot be performed with a single PHPO, unlike the adiabatic system case for which a single PPO was sufficient. Therefore, we must determine another approach through which summation of all the PHPOs (as quantum interferences between PHPOs) can be correctly performed. In addition, note also that there is a countably infinite number of PHPOs, owing to the infinite number of possible hopping trajectories. 

To enable the summation of all the PHPOs in Eq.~(\ref{eq:NonAdiTrace01}), we introduce a set, $\mathscr{S}$, for each adiabatic surface. This $\mathscr{S}$ is defined as a set of PHPOs that (i) pass through the same phase space point and (ii) cannot be factorized as a combination of other PHPOs in $\mathscr{S}$; that is, any pair of PHPOs in $\mathscr{S}$ is coprime: $\{ ^{\forall} \Gamma, ^{\forall}\Gamma' \in \mathscr{S} | \  \Gamma' \not\subset \Gamma \ \vee \ \Gamma\setminus\Gamma' \notin \mathscr{S} \}$. Hereafter, the elements of  $\mathscr{S}$ will be referred to as prime PHPOs. Rewriting Eq.~(\ref{eq:NonAdiTrace01}) in terms of $\mathscr{S}$ leads to
\begin{eqnarray}
\Omega(E) &\propto&\sum_{\mathscr{S}_i \in \{\mathscr{S}\}} \Omega_{\mathscr{S}_i}(E),
\label{eq:NonAdiTrace01.6}\\
\Omega_{\mathscr{S}_i}(E)&\equiv &  \sum_{k=0}^{\infty} \left\{\sum_{\lambda \in \mathscr{S}_i} \mathscr{G}_{\lambda} \right\}^k = \frac{1 }{1-\left\{\sum_{\lambda \in \mathscr{S}_i}\mathscr{G}_{\lambda} \right\}},
\label{eq:NonAdiTrace01.7}
\end{eqnarray}
where $\{\mathscr{S}\}$ is a set of sets. The element, $\mathscr{S}_i$, is the set of prime PHPOs that begin and end at a point on the $i$-th adiabatic surface. So, the number of elements of $\{\mathscr{S}\}$ equals the number of adiabatic surfaces considered. In Eq.~(\ref{eq:NonAdiTrace01.7}), all the PHPO are included as a combination of the prime PHPOs in $\mathscr{S}_i$. Especially, all the prime PHPOs in each $\mathscr{S}_i$ for one-dimensional systems can be written finitely (as explained concretely in Sec.~\ref{sec:ex}) and, therefore, summation over $\lambda$ can be performed. Thus, we conclude from Eq.~(\ref{eq:NonAdiTrace01.7}) that the semiclassical quantization condition for one-dimensional nonadiabatic systems is 
\begin{eqnarray}
\sum_{\lambda \in {\rm prime\ PHPOs}} \mathscr{G}_{\lambda}  = 1.
\label{eq:NonAdiTrace02}
\end{eqnarray}

\section{Concrete example \label{sec:ex}}
\subsection{Model}
\begin{figure}
\includegraphics[width=\linewidth]{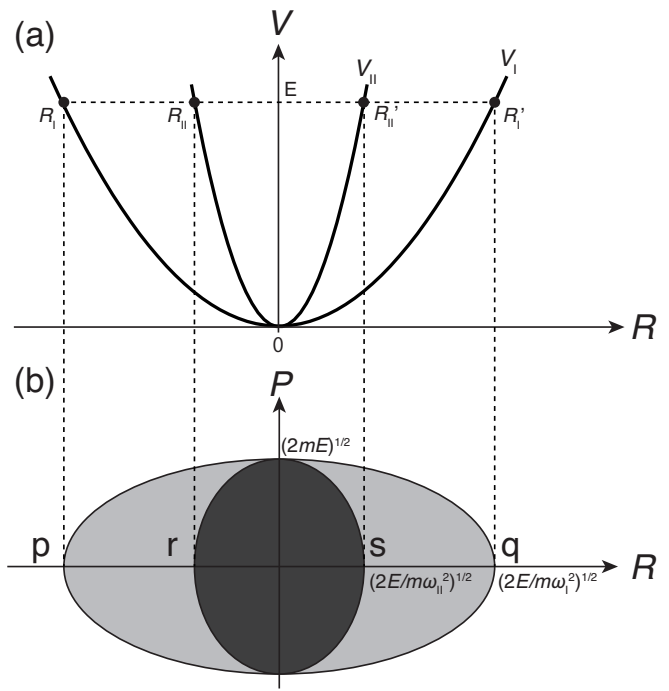}
\caption{(a) Two adiabatic harmonic surfaces. These adiabatic surfaces interact nonadiabatically at the origin only. (b) Phase space area enclosed by classical trajectories on $V_{\rm I}$ and $V_{\rm II}$ are represented by light and dark gray ellipses, respectively. The areas of the former and latter are referred to as $S_{\rm I}^{\rm cl}$ and $S_{\rm II}^{\rm cl}$, respectively, in the text.
\label{fig:DoubleHarmPhaseTraj}}
\end{figure}
We apply the semiclassical quantization presented here to a simple nonadiabatic model consisting of two adiabatic quantum harmonic oscillators, $V_{\rm I}(R)$ and $V_{\rm II}(R)$, as shown in Fig.~\ref{fig:DoubleHarmPhaseTraj}(a). We have
\begin{eqnarray}
V_{\rm I}(R) &=& \frac{1}{2}M \omega_{\rm I}^2 R^2, 
\label{eq:App01} 
\end{eqnarray}
and
\begin{eqnarray}
V_{\rm II}(R) &=& \frac{1}{2}M \omega_{\rm II}^2 R^2,
\label{eq:App02} 
\end{eqnarray}
where $\omega_{\rm I} = 27.6$ [${\rm kcal^{1/2} mol^{-1/2} \AA^{-1} amu^{-1/2}}$], $\omega_{\rm II} = 38.64$ [${\rm kcal^{1/2} mol^{-1/2} \AA^{-1} amu^{-1/2}}$], and $M$ is 1 [amu]. The eigenstates corresponding to these two adiabatic surfaces are represented as $|{\rm I};R \rangle$ and $|{\rm II};R \rangle$, respectively, and, to induce nonadiabatic transitions, these adiabatic eigenstates are defined by a set of diabatic states, $|\Psi_{\rm A} \rangle$ and $|\Psi_{\rm B} \rangle$, such that
\begin{eqnarray}
\left(
    \begin{array}{c}
      |{\rm I};R \rangle \\
      |{\rm II};R \rangle
     \end{array} 
 \right)
 &\equiv&
\left(
    \begin{array}{cc}
      \cos\phi (R) & \sin\phi (R)\\
      -\sin\phi (R)& \cos\phi (R)
     \end{array} 
 \right)
  \left(
      \begin{array}{c}
        |\Psi_{\rm A}\rangle \\
        |\Psi_{\rm B}\rangle
       \end{array} 
  \right),
\label{eq:App03} \\
\phi(R) &\equiv& \phi \theta(R),
\label{eq:App03.5}
\end{eqnarray}
where $\theta(R)$ is the Heaviside step function. The parameter $\phi\ (0\le \phi \le \pi/2)$ sets the nonadiabaticity in this model, and $\phi = \pi/2$ and $0$ correspond to the diabatic and adiabatic limits, respectively. 
The overlap integrals of the adiabatic eigenstates, which are indexes of nonadiabaticity in the path integral picture, can then be calculated as
\begin{eqnarray}
&&\left(
    \begin{array}{cc}
      \langle {\rm I};R'|{\rm I};R\rangle  & \langle {\rm I};R'  | {\rm II};R \rangle  \\
      \langle {\rm II};R'|  {\rm I};R\rangle  & \langle {\rm II};R'  | {\rm II};R \rangle 
     \end{array} 
 \right) \nonumber \\
&&\hspace{5mm}=
  \left\lbrace 
  \begin{array}{l}
  \left(
      \begin{array}{cc}
        \cos\phi & -\sin\phi \\
        \sin\phi & \cos\phi 
       \end{array} 
   \right) 
   \ {\rm for}\ R'<0 \land 0\le R, \\ 
    \left(
        \begin{array}{cc}
          1 & 0 \\
          0 & 1 
         \end{array} 
     \right)
     \hspace{15.5mm} {\rm otherwise}.
  \end{array} 
  \right.
  \label{eq:App06} 
\end{eqnarray}
The first-order derivative couplings calculated from Eq.~(\ref{eq:App03}) diverge at the origin of the $R$-axis only:
$\langle {\rm I};R|\frac{d}{dR} |{\rm II};R\rangle = - \langle {\rm II};R|\frac{d}{dR} |{\rm I};R\rangle = -\phi \delta(R)$. Therefore, wave packets can nonadiabatically transfer to another adiabatic surface only when they pass through the origin of the $R$-axis. In semiclassical treatments, the trajectories can hop to another adiabatic surface only when they pass through the origin.

\subsection{Bit sequences representing nonadiabatic dynamics\label{subsec:bit}}
\begin{figure}
\includegraphics[width=\linewidth]{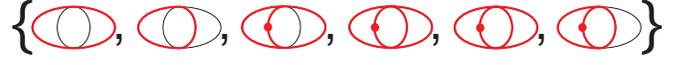}
\caption{(color online).  Prime PHPOs in $\mathscr{S}_{\rm I}$ [Eq.~(\ref{eq:App07.5})]. Each prime PHPO is graphically represented with a red thick curve and dot. The dot represents the bold number  ${\bf 1}$.  These graphical orbits represent $\dot{0} \dot{0}$, $\dot{0} \dot{1}$, $\dot{0} 0 {\bf 1} \dot{0}$, $\dot{0} 0 {\bf 1} \dot{1}$, $\dot{0} 1 {\bf 1} \dot{0}$, and $\dot{0} 1 {\bf 1} \dot{1}$ from left, respectively. Note that the geometries of $\dot{0}0{\bf 1}\dot{1}$ and $\dot{0}1{\bf 1}\dot{0}$ (4-th and 5-th from left, respectively) are identical.
\label{fig:coprimePHPOs}}
\end{figure}

Let us concretely describe the set ($\mathscr{S}_{\rm I}$) of prime PHPOs. First of all, we consider PHPOs that begin at the phase space point, ``p'', in Fig.~\ref{fig:DoubleHarmPhaseTraj}(b), where 
\begin{eqnarray}
    \begin{array}{l}
     p \rightarrow q \rightarrow p,\\
     p \rightarrow s \rightarrow  p, \\
     p \rightarrow q \rightarrow p \rightarrow s \rightarrow  p, \\
     p \rightarrow q \rightarrow r \rightarrow s \rightarrow  p, \\
     \hspace{20pt}\vdots\hspace{65pt}.
     \end{array} 
\label{eq:App07.1} 
\end{eqnarray}
To simplify the representation, we assign a bit sequence to each PHPO such that ``0" (``1") indicates that a PHPO has passed through  ``p" or ``q" (``r" or ``s"). The PHPOs in Eq.~(\ref{eq:App07.1}) are then denoted by
\begin{eqnarray}
\dot{0} \dot{0},\ \dot{0} \dot{1},\ \dot{0} 0 0 \dot{1},\ \dot{0} 0 1 \dot{1},\ \ldots ,
\label{eq:App07.2} 
\end{eqnarray}
where the dots indicate the start and end points of one cycle. When these cyclic bit sequences are considered as fractional parts of real numbers, all the cyclic bits correspond to a subset of rational numbers between 0 and 1/2. Second, to ensure that any pair of PHPOs in $\mathscr{S}_{\rm I}$ is coprime, we must exclude abundant PHPOs. For example, $\dot{0} 0 0 \dot{1}$ is not coprime with $\dot{0} \dot{0}$ and $\dot{0} \dot{1}$ because $\dot{0} 0 0 \dot{1}$ is a combination of $00$ and $01$. We can thus exclude abundant PHPOs by retaining only bit sequences in which odd-numbered bits, apart from the first bit, are 1. We denote the bit sequences in which all odd-numbered bits are 1 (e.g., 1, 101, 111, 10101, $\cdots$, $1011101\cdots 1$, etc.) as ${\bf 1}$.  Finally, all the prime PHPOs in $\mathscr{S}_{\rm I}$ can be listed as
\begin{eqnarray}
\mathscr{S}_{\rm I}=
\{
         \dot{0} \dot{0},\ 
         \dot{0} \dot{1},\ 
         \dot{0} 0 {\bf 1} \dot{0},\ 
         \dot{0} 0 {\bf 1} \dot{1},\ 
         \dot{0} 1 {\bf 1} \dot{0},\ 
         \dot{0} 1 {\bf 1} \dot{1}
\}.
\label{eq:App07.5} 
\end{eqnarray}
The prime PHPOs in $\mathscr{S}_{\rm I}$ are shown graphically as trajectories in the phase space of slow DOF in Fig.~\ref{fig:coprimePHPOs}. These graphical orbits show transitions between two periodic orbits of adiabatic harmonic oscillators. Note that, in the adiabatic (diabatic) limit, only $\dot{0} \dot{0}$ ($\dot{0} \dot{1}$) is used, because the overlap integrals of other prime PHPOs vanish (see Eqs. (\ref{eq:App08.5}) -- (\ref{eq:App09})).

\subsection{Semiclassical quantization\label{subsec:semi_quan}}
To calculate $\mathscr{G}$ [Eq.~(\ref{eq:NonAdiTrace01.3})], all the geometrical quantities of the prime PHPOs are calculated in this subsection. We first note from Fig.~\ref{fig:coprimePHPOs} that the geometries of $\dot{0}01\dot{1}$ and $\dot{0}11\dot{0}$ are identical, and therefore  $S_{\dot{0} 0 1 \dot{1}}^{\rm cl}=S_{\dot{0} 1 1 \dot{0}}^{\rm cl},\ \xi_{\dot{0} 0 1 \dot{1}}=\xi_{\dot{0} 1 1 \dot{0}}$, and  $\nu_{\dot{0} 0 1 \dot{1}}=\nu_{\dot{0} 1 1 \dot{0}}$. 
From Fig.~\ref{fig:DoubleHarmPhaseTraj}(b), in which the areas of the light and dark gray ellipses are $S^{\rm cl}_{\rm I}$ and $S^{\rm cl}_{\rm II}$, respectively, we find that 
\begin{eqnarray}
	S_{\dot{0} \dot{0}}^{\rm cl} &=& S^{\rm cl}_{\rm I},\\  
	S_{\dot{0} \dot{1}}^{\rm cl} &=& \frac{1}{2}S^{\rm cl}_{\rm I}+\frac{1}{2}S^{\rm cl}_{\rm II},\\  
	S_{\dot{0} 0 1 \dot{0}}^{\rm cl} &=& \frac{3}{2}S^{\rm cl}_{\rm I}+\frac{1}{2}S^{\rm cl}_{\rm II} ,\\
	S_{\dot{0} 0 1 \dot{1}}^{\rm cl} &=& S^{\rm cl}_{\rm I}+S^{\rm cl}_{\rm II} ,\\
	S_{\dot{0} 1 1 \dot{1}}^{\rm cl} &=& \frac{1}{2}S^{\rm cl}_{\rm I}+\frac{3}{2}S^{\rm cl}_{\rm II},
\label{eq:App08} 
\end{eqnarray}
where $S^{\rm cl}_{\rm I} = 2\pi E/\omega_{\rm I}$  and $S^{\rm cl}_{\rm II} = 2\pi E/\omega_{\rm II}$. In calculating $\xi$, we use the overlap integrals defined in Eq.~(\ref{eq:App06}) when the prime PHPO passes through the origin of the $R$-axis to give 
\begin{eqnarray}
\xi_{\dot{0} \dot{0}} &=& \cos^2 \phi, 
\label{eq:App08.5} \\
\xi_{\dot{0} \dot{1}} &=& \sin^2 \phi, \\
\xi_{\dot{0} 0 1 \dot{0}} &=& \cos^2 \phi\sin^2 \phi, \\
\xi_{\dot{0} 0 1 \dot{1}} &=& -\cos^2 \phi\sin^2 \phi, \\
\xi_{\dot{0} 1 1 \dot{1}} &=& \cos^2 \phi\sin^2 \phi.
\label{eq:App09} 
\end{eqnarray}
The Maslov indexes for these cycles are $\nu_{\dot{0} \dot{0}}     =  \nu_{\dot{0} \dot{1}} = 2$ and $\nu_{\dot{0} 01 \dot{0}} =  \nu_{\dot{0} 01 \dot{1}} = \nu_{\dot{0} 11 \dot{1}} = 4$.
We can now calculate $\mathscr{G}_{\dot{0} \dot{0}},\ \mathscr{G}_{\dot{0} \dot{1}},\ \mathscr{G}_{\dot{0} 0 1 \dot{0}},\ \mathscr{G}_{\dot{0} 0 1 \dot{1}},\  \mathscr{G}_{\dot{0} 1 1 \dot{0}}$, and $\mathscr{G}_{\dot{0} 1 1 \dot{1}}$, but we have not yet considered the bit sequences represented by ${\bf 1}$. As these bit sequences include all combinations of $10$ and $11$, the contribution to the DOS can be summarized as
\begin{eqnarray}
\Omega_{\bf 1} \equiv \sum_{k=0}^{\infty} \left( \mathscr{G}_{\dot{1}\dot{0}} + \mathscr{G}_{\dot{1}\dot{1}}  \right)^k = \frac{1}{1-\left( \mathscr{G}_{\dot{1}\dot{0}} + \mathscr{G}_{\dot{1}\dot{1}}  \right)},
\label{eq:App13}
\end{eqnarray}
where $\mathscr{G}_{\dot{1}\dot{0}}$ and $\mathscr{G}_{\dot{1}\dot{1}}$ can be calculated in a similar manner to $\mathscr{G}_{\dot{0}\dot{1}}$ and $\mathscr{G}_{\dot{0}\dot{0}}$, respectively, by exchanging $\omega_{\rm I}$ and $\omega_{\rm II}$. We then find that
\begin{eqnarray}
\mathscr{G}_{\dot{0} 0 {\bf 1} \dot{0}}  = \mathscr{G}_{\dot{0} 0 1 \dot{0}} \Omega_{\bf 1},\ 
\mathscr{G}_{\dot{0} 0 {\bf 1} \dot{1}}  = \mathscr{G}_{\dot{0} 0 1 \dot{1}} \Omega_{\bf 1},\nonumber \\
\mathscr{G}_{\dot{0} 1 {\bf 1} \dot{0}}  = \mathscr{G}_{\dot{0} 1 1 \dot{0}} \Omega_{\bf 1},\ 
\mathscr{G}_{\dot{0} 1 {\bf 1} \dot{1}}  = \mathscr{G}_{\dot{0} 1 1 \dot{1}} \Omega_{\bf 1}.
\label{eq:App14}
\end{eqnarray}
We can now perform an analytical calculation of the DOS (Eq.~(\ref{eq:NonAdiTrace01.7})) and the semiclassical quantization condition (Eq.~(\ref{eq:NonAdiTrace02})) for $\mathscr{S}_{\rm I}$. 

To calculate the total DOS (Eq.~(\ref{eq:NonAdiTrace01.6})), a set, $\mathscr{S}_{\rm II}$, and the corresponding geometrical quantities for each prime PHPO in $\mathscr{S}_{\rm II}$ are also required. All the prime PHPOs in $\mathscr{S}_{\rm II}$, using the same arguments starting from ``r" in Fig.~\ref{fig:DoubleHarmPhaseTraj} (b), can be expressed as 
\begin{eqnarray}
\mathscr{S}_{\rm II}=
\{
         \dot{1} \dot{1},\ 
         \dot{1} \dot{0},\ 
         \dot{1} 1 {\bf 0} \dot{1},\ 
         \dot{1} 1 {\bf 0} \dot{0},\ 
         \dot{1} 0 {\bf 0} \dot{1},\ 
         \dot{1} 0 {\bf 0} \dot{0}
\},
\label{eq:App15} 
\end{eqnarray}
where ${\bf 0}$ represents bit sequences in which all odd-numbered bits are 0. The geometrical quantities for $\mathscr{S}_{\rm II}$ can be then calculated in the same way as those for $\mathscr{S}_{\rm I}$. Thus, these two sets and the corresponding geometrical quantities enable an analytical calculation of the total DOS in Eq.~(\ref{eq:NonAdiTrace01.6}).

\begin{figure}
\includegraphics[width=.8\linewidth]{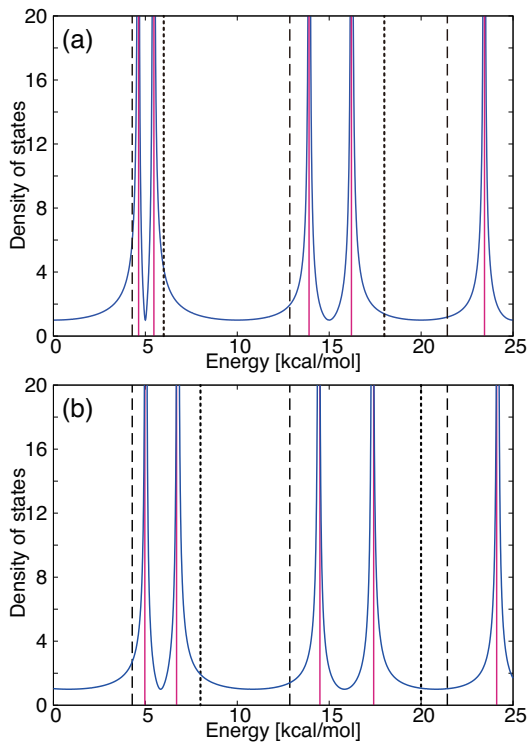}
\caption{(color online). Absolute value of DOSs. Blue solid curve and red solid line represent DOSs calculated using the semiclassical and the numerically exact nonadiabatic quantizations, respectively. Black broken and dotted lines represent DOSs calculated using the exact adiabatic quantizations on $V_{\rm I}$ and $V_{\rm II}$, respectively. (a) $V_{\rm I}$ and $V_{\rm II}$ are degenerate at the origin, [Eqs.~(\ref{eq:App01}) and (\ref{eq:App02})]. (b) $V_{\rm I}$ and $V_{\rm II}$ are non-degenerate, [Eqs.~(\ref{eq:App01}) and (\ref{eq:App16}) with $\Delta E=2.0$ [kcal/mol]].
\label{fig:adi_nonadi_exact_semicl_phi-0.33PI}}
\end{figure}

Figure \ref{fig:adi_nonadi_exact_semicl_phi-0.33PI} (a) compares four DOSs for the case of $\phi=\pi/3$. The first DOS (blue solid curve) is calculated using the present semiclassical quantization for nonadiabatic systems, Eq.~(\ref{eq:NonAdiTrace01.6}), the second (red solid line) is determined using the numerically exact method for nonadiabatic systems, while the third and fourth (black broken and dotted lines, respectively) are adiabatic DOSs calculated analytically on each adiabatic harmonic potential. So, the last two DOSs do not contain any nonadiabatic effects. Note that detail of how to numerically calculate the nonadiabatic exact quantum energy levels is given in Ref.~\onlinecite{M.Fujii2011}. The numerical parameters used in the present exact calculation are $J=129$, $R_{min}=-3.0 [{\rm \AA}]$, and $R_{max}=3.0 [{\rm \AA}]$. Because the energy levels  of quantum steady states are indicated by divergences in the DOS, we see that the energy levels determined by the present semiclassical quantization agree with the numerically exact nonadiabatic energy levels, rather than the adiabatic energy levels, very well. We also consider another case in which the two harmonic potentials are non-degenerate at the origin. For this non-degenerate case, the second harmonic potential, Eq.~(\ref{eq:App02}), is replaced with 
\begin{eqnarray}
V_{\rm II}(R) &=& \frac{1}{2}M \omega_{\rm II}^2 R^2 + \Delta E.
\label{eq:App16} 
\end{eqnarray}
Therefore, the first integral in Eq.~(\ref{eq:AdiTrace03}) should be used to calculate the classical action integrals because, although the energies of the PHPOs are not conserved, their momenta are conserved at each hopping\footnote{Note that the momentum and energy conservation are
equivalent for the present case because of the particularity of the harmonic potentials.}. Figure~\ref{fig:adi_nonadi_exact_semicl_phi-0.33PI} (b) compares the four DOSs (similar to Fig.~\ref{fig:adi_nonadi_exact_semicl_phi-0.33PI} (a) for the non-degenerate model), with $\Delta E=2.0$ [kcal/mol]. One can confirm that, even in the non-degenerate case, the energy levels calculated with the present semiclassical quantization for the nonadiabatic systems agree with the numerically exact nonadiabatic energy levels.

\subsection{Chaotic dynamics in nonadiabatic systems \label{subsec:chaos}}
\begin{figure}
\includegraphics[width=\linewidth]{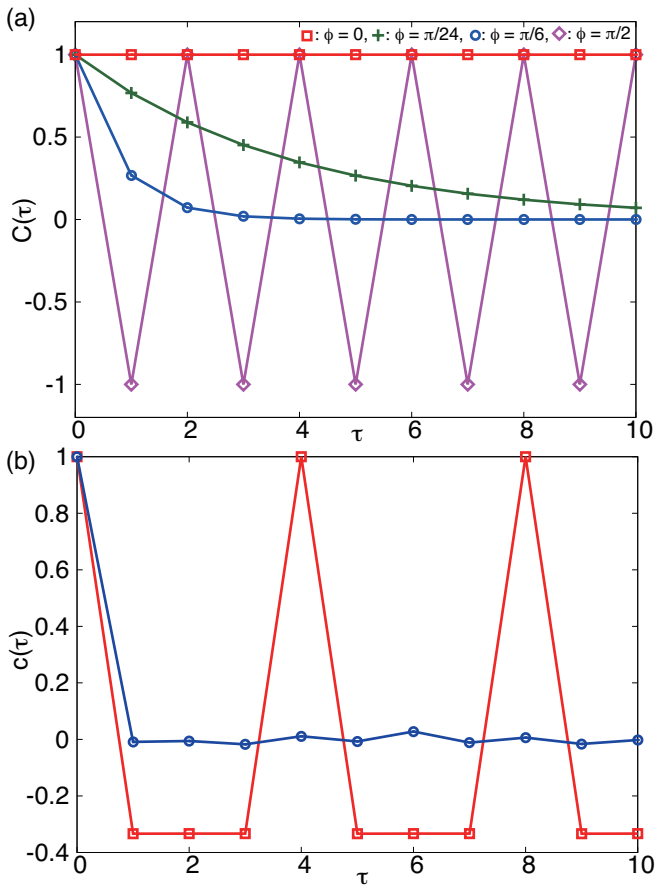}
\caption{(color online) (a) Autocorrelation functions [Eq.~(\ref{eq:chaos01})] for four cases: $\phi=0$ (square), $\phi=\pi/24$ (cross), $\phi=\pi/6$ (circle), and $\phi=\pi/2$ (diamond). (b)  Autocorrelation functions [Eq.~(\ref{eq:chaos06})] for $\dot{0}01\dot{0}$ (square) and the binary number of the decimal part of $\pi$ (circle).
\label{fig:autocorrelations}}
\end{figure}
Chaotic dynamics induced by nonadiabatic transitions is one of the most interesting topics of chemical dynamics, and is occasionally studied as a topic in quantum chaos~\cite{A.Bulgac1995,H.Fujisaki2001}. The chaotic dynamics in nonadiabatic systems can be investigated using the classical electron analog model and interpreted as the collapse of tori in the total phase space of the electron analog and nuclei~\cite{R.L.Whetten1985}. Returning to classical dynamics, the physical origin of classical chaotic dynamics in nonadiabatic systems is obviously the interaction between the fast and slow DOFs. This interaction also leads to stochastic dynamics known as Arnold diffusion along the resonance lines of the fast and slow DOFs~\cite{S.Fuchigami2003}. Based on correspondence with classical chaotic dynamics, it is apparent that the quantum eigenvalues of any variable consisting of only the fast DOF cannot be good quantum numbers. Therefore, nonadiabatic transitions can occur. In addition to the above discussion concerning the relationship of the nonadiabatic transition and chaotic motion in the classical limit, in this section, we present another discussion from the semiclassical perspective. Here, the chaotic motion embedded in the nonadiabatic dynamics is investigated using the symbolic dynamics introduced in Subsec.~\ref{subsec:bit}.

An autocorrelation function for the symbolic dynamics
\begin{eqnarray}
C(\tau) \equiv  \left\langle\left\langle  \left( s_i -\bar{s}_d \right) \left( s_{i+\tau} -\bar{s}_d \right) \right\rangle\right\rangle / \sigma_d^2,
\label{eq:chaos01}
\end{eqnarray}
is introduced to clarify the chaotic motion in the classical limit. This autocorrelation function decays as time progresses, if the nonadiabatic dynamics represented by the symbolic dynamics involve chaotic motion. In Eq.~(\ref{eq:chaos01}),  $s_i$ represents the $i$-th bit variable in a bit sequence, e.g., $s_i=0$ or $1$. The double bra-ket $\langle\langle \cdots  \rangle \rangle$ represents double averages: the first is the average in time evolution for a bit sequence, while the second is the average of all bit sequences. Therefore, the average and variance of each bit variable, respectively, follow 
\begin{eqnarray}
\bar{s}_d &\equiv& \langle\langle s_i \rangle \rangle= \frac{1}{2},
\label{eq:chaos02} \\
\sigma^2_d &\equiv& \langle  \langle(s_i -  \bar{s})^2 \rangle\rangle = \frac{1}{4}.
\label{eq:chaos03}
\end{eqnarray}
In the calculation of this autocorrelation function, the symbolic dynamics can be treated as stochastic bit sequences in which the bits are inverted with a probability that is equal to the probability of semiclassical hopping. This is because all bit sequences are considered in the averaging. Then, the probability of bit inversion for the model we consider here is expressed as\footnote{See also Eq.~(41) in Ref.~\onlinecite{M.Fujii2011}.}
\begin{eqnarray}
p_{\rm inv} = \frac{\sin\phi}{\sin\phi + \cos\phi}.
\label{eq:chaos04}
\end{eqnarray} 
The autocorrelation function can be analytically calculated as 
\begin{eqnarray}
C(\tau) 
&=& 2\sum_{k=0}^{2k \leq \tau}
\left(
    \begin{array}{c}
      \tau \\
      2k
     \end{array} 
 \right)
 p_{\rm inv}^{2k}\left( 1- p_{\rm inv} \right)^{\tau -2k} - 1, \nonumber \\
&=& \left( 1- 2p_{\rm inv} \right) ^{\tau} ,
\label{eq:chaos05}
\end{eqnarray}
 and the autocorrelation functions for four cases ($\phi=0$, $\phi=\pi/24$, $\phi=\pi/6$, and $\phi=\pi/2$) are shown in Fig.~\ref{fig:autocorrelations} (a). The autocorrelation functions in the adiabatic ($\phi=0$) and the diabatic limit ($\phi=\pi/2$) do not decay because their symbolic trajectories are periodic. 
The symbolic trajectories for the adiabatic limit are $\dot{0}\dot{0}$ and $\dot{1}\dot{1}$, because the semiclassical trajectories never hop, while the symbolic trajectories for the diabatic limit are $\dot{0}\dot{1}$ and $\dot{1}\dot{0}$, because the semiclassical trajectories must hop between adiabatic harmonic potentials at the origin. 
On the other hand, in the nonadiabatic cases ($\phi=\pi/24$ and $\pi/6$) the autocorrelation functions decay. In particular, the autocorrelation function in the strong nonadiabatic case ($\phi=\pi/6$) decays more rapidly than that of the weak nonadiabatic case ($\phi=\pi/24$). Therefore, by considering the decay of the autocorrelation functions, we can confirm that chaotic dynamics appear at the classical limit of the nonadiabatic dynamics.

In contrast to the semiclassical analysis of the decay of the autocorrelation function, quantum autocorrelation functions for isolated quantum systems do not decay, even if the system is nonadiabatic. To clarify this difference, we analyze an autocorrelation function for a bit sequence
\begin{eqnarray}
c(\tau) \equiv  \left\langle  \left( s_i -\bar{s}_s \right) \left( s_{i+\tau} -\bar{s}_s \right) \right\rangle / \sigma_s^2,
\label{eq:chaos06}
\end{eqnarray}
with $\bar{s}_s \equiv \left\langle  s_i \right\rangle$ and $\sigma_s^2 \equiv  \langle \left( s_i -\bar{s}_s \right)^2 \rangle$, where only the average time evolution of a bit sequence is taken into consideration. In Fig.~\ref{fig:autocorrelations} (b), two autocorrelations are shown: the first is an autocorrelation function for $\dot{0}01\dot{0}$ that was used in the semiclassical quantization, while the second is an autocorrelation function for the binary number of the decimal part of $\pi$. In the calculation of the latter, only the first 3,000 binary bits ($001001000011\cdots$) were used. We can confirm from Fig.~\ref{fig:autocorrelations} (b) that the autocorrelation functions for the periodic bit sequences corresponding to the rational numbers do not decay, while those for the non-periodic bit sequences corresponding to the irrational numbers decay rapidly. In Subsec.~\ref{subsec:semi_quan}, we demonstrated that the semiclassical quantization of nonadiabatic systems can be conducted using only PHPOs corresponding to the periodic bit sequences. Namely, the non-periodic orbits causing the decay of the autocorrelation function decay are not required for the quantization of the nonadiabatic systems and, hence, the quantum autocorrelation functions do not decay.

\section{Concluding remarks\label{sec:concluding}}
We have presented a new derivation of the Schr\"{o}dinger equation holding the nonadiabatic couplings from the nonadiabatic path integral and a nonadiabatic trace formula leading to a semiclassical quantization condition for one-dimensional nonadiabatic systems. The present approach shows that the quantization mechanism is based on a set of prime PHPOs, $\mathscr{S}$, in which all prime PHPOs pass through the same phase space point and any pair of prime PHPOs is coprime.  Specifically, $\mathscr{S}$, which is invariant under time evolution in the phase space, corresponds to the quantum eigenstates in the classical limit. This semiclassical quantization was applied to a simple nonadiabatic model and accurately reproduced the exact quantum energy levels. In addition, the chaotic dynamics embedded in the nonadiabatic dynamics were also shown using symbolic dynamics. In summary, nonadiabatic (hopping) orbits can be categorized to two classes of dynamics using the symbolic dynamics. The first class is a set of hopping periodic orbits which correspond to rational numbers represented as periodic binary bits, while the second class is a set of chaotic orbits which correspond to irrational numbers represented as non-periodic binary bits. The hopping periodic orbits in the first class contribute to the semiclassical quantization, while the hopping orbits in the second class do not contribute and cause a decay in the autocorrelation functions in the classical limit.
The present paper has shed new light on nonadiabatic phenomena. However, a number of issues remain, e.g., multidimensional problems such as conical intersections \cite{S.Matsika2011}, quantitative analyses of the violation of the quantum adiabatic theorem~\cite{K.-P.Marzlin2004,D.M.Tong2005,*D.M.Tong2007,M.H.S.Amin2009,J.Ortigoso2012}, and the relationship with the Riemann hypothesis~\cite{M.V.Berry1990,H.M.Edwards1974,M.V.Berry1988}. In the remainder of this section, we discuss the outstanding issues which should be addressed in the future. 

Extending the semiclassical quantization of nonadiabatic systems to that of multidimensional systems seems complicated, although extending the nonadiabatic path integral and the semiclassical kernel to multidimensional systems is straightforward. Even in adiabatic cases, the semiclassical quantization of multidimensional systems is quite different to that of one dimensional systems. According to the EBK semiclassical quantization~\cite{J.B.Keller1958}, tori in the phase space are quantized for multidimensional systems, in contrast to the semiclassical quantization of primitive periodic orbits, which are obvious solutions of classical mechanics in a one-dimensional system. In other words, independent periodic paths on the tori, which are not necessarily classical trajectories, are quantized. Considering this situation, how can we correctly formulate nonadiabatic interference between the tori in a similar manner to the decomposition of the PHPOs into combinations of prime PHPOs in one-dimensional systems? In chaotic energy regions in which the tori break, could periodic orbits on different adiabatic surfaces pass through the same phase space point? In addition to these remaining questions, the appropriate treatment of conical intersections is also a challenging issue. In cases for which conical intersections exist, hopping paths that hop with conservation of both momentum and energy at the conical intersections will mainly contribute to the semiclassical quantization. As we have stated in this paragraph, the semiclassical quantization of nonadiabatic multidimensional systems requires further investigation. 

The adiabatic theorem~\cite{P.Ehrenfest1916,M.Born1928,T.Kato1950,A.Messiah1962} is one of the most common theorems with wide applications in various scientific fields. 
However, some years ago, Marzlin and Sanders demonstrated an inconsistency as regards the applicability of the quantum adiabatic theorem~\cite{K.-P.Marzlin2004}. 
This theorem has recently been applied in quantum computing~\cite{E.Farhi2001} and its violation is a source of errors in quantum computing. Therefore, the quantum adiabatic theorem has very recently been intensively re-investigated to determine necessary and sufficient conditions~\cite{D.M.Tong2005,D.M.Tong2007,M.H.S.Amin2009,J.Ortigoso2012}. Occasionally, the Landau--Zener~\cite{L.D.Landau1932,C.Zener1932} formula can be used in quantitative analyses of the violation of the quantum adiabatic theorem\cite{G.E.Santoro2005,S.Suzuki2005} but, in these analyses, the quantum effects of the slow variable that is usually treated as a parametric degree of freedom are not considered. The present nonadiabatic path integral helps us to investigate quantum effects in violation of the quantum adiabatic theorem. 

The relationship between a distribution of nontrivial zeros in the zeta-function and that of eigenvalues of the quantum Hamiltonians of chaotic systems is well-known as a unique result~\cite{M.V.Berry1990,M.V.Berry1988}. In addition, chaotic dynamics due to nonadiabatic transitions are an interesting phenomenon known as quantum chaos~\cite{A.Bulgac1995,H.Fujisaki2001,R.L.Whetten1985}. Thus, we can easily imagine that the distribution of nontrivial zeros in the zeta-function and that of eigenstates of nonadiabatic systems are related to each other. To discuss this relationship that sheds new light on the Riemann hypothesis~\cite{H.M.Edwards1974}, the nonadiabatic trace formula and chaotic dynamics represented by symbolic dynamics are a suitable starting point. In particular, the prime PHPOs that can generate all PHPOs seem to resemble prime numbers, which generate all natural numbers. At present, we have not found a clear guiding principle for this relationship, but it merits consideration based on the nonadiabatic path integral and trace formula.

To conclude the present paper, we have discussed some theoretical matters that could potentially be explored subsequent to this study. In addition to these theoretical matters, the various potential applications of the nonadiabatic path integral to realistic molecules are also interesting and important areas of investigation, e.g., in surface scattering, photoisomerization in vision, control of molecular dynamics, quantum computing, and organic solar cells. To this end, new ideas, concepts, and techniques must be established that will contribute to deeper understanding of nonadiabatic transitions in the future.

\section*{Acknowledgement}
MF appreciates valuable discussions with Prof. H. Ushiyama concerning symbolic dynamics, and with Prof. O. K{\"u}hn regarding future applications of the present theories. This work was supported by JSPS KAKENHI Grant No.~24750012 and JST, CREST.

\providecommand{\noopsort}[1]{}\providecommand{\singleletter}[1]{#1}%

\end{document}